\documentclass[12pt]{article}
\usepackage{graphicx}
\begin{document}

\bigskip
\centerline{\bf \large Child mortality in Penna ageing model}

\bigskip

\noindent
D. Stauffer$^{1,2}$ and S. Moss de Oliveira $^{1,3}$

\bigskip
\noindent
$^1$ Laboratoire PMMH, \'Ecole Sup\'erieure de Physique et de Chimie
Industrielles, 10 rue Vauquelin, F-75231 Paris, France

\medskip
\noindent
$^2$ Visiting from Inst. for Theoretical Physics, Cologne University, 
D-50923 K\"oln, Euroland

\medskip
\noindent
$^3$ Visiting from Instituto de F\'{\i}sica, Universidade
Federal Fluminense; Av. Litor\^{a}nea s/n, Boa Viagem,
Niter\'{o}i 24210-340, RJ, Brazil

\bigskip

\noindent
{\bf
Assuming the deleterious mutations in the Penna ageing model to affect
mainly the young ages, we get an enhanced mortality at very young age,
followed by a minimum of the mortality, and then the usual exponential
increase of mortality with age.
} 

\bigskip

The standard Penna ageing model, based on strings of 32 bits for each 
individual, gives good agreement with human mortality at middle and 
old age $a$, where the mortality function $\mu(a)$ inreases exponentially
with age (Gompertz law)\cite{penna,book,newbook}. In childhood, however, 
human mortality has a 
minimum at about 7 to 10 years, preceded by a higher mortality at younger
ages from childhood diseases. This mortality minimum was previously found
in the Penna model by introducing housekeeping genes \cite{cebrat}
or different time scales \cite{magdon}. We now offer a simpler modification
giving this minimum.

Children are more susceptible to disease and may suffer from deadly genetic
defects. This is ignored in the standard Penna model where each bit position
(corresponding to the age at which a life-treatening disease starts) is
mutated with equal probability. Now instead we assume a mutation probability
inversely proportional to the bit position = age. Otherwise we use the
standard parameters as in the program published in \cite{book}; in particular,
one mutation happens at birth, and three active diseases kill. Fig. 1 shows
the desired mortality minimum.

\begin{figure}[hbt]
\begin{center}
\includegraphics[angle=-90,scale=0.5]{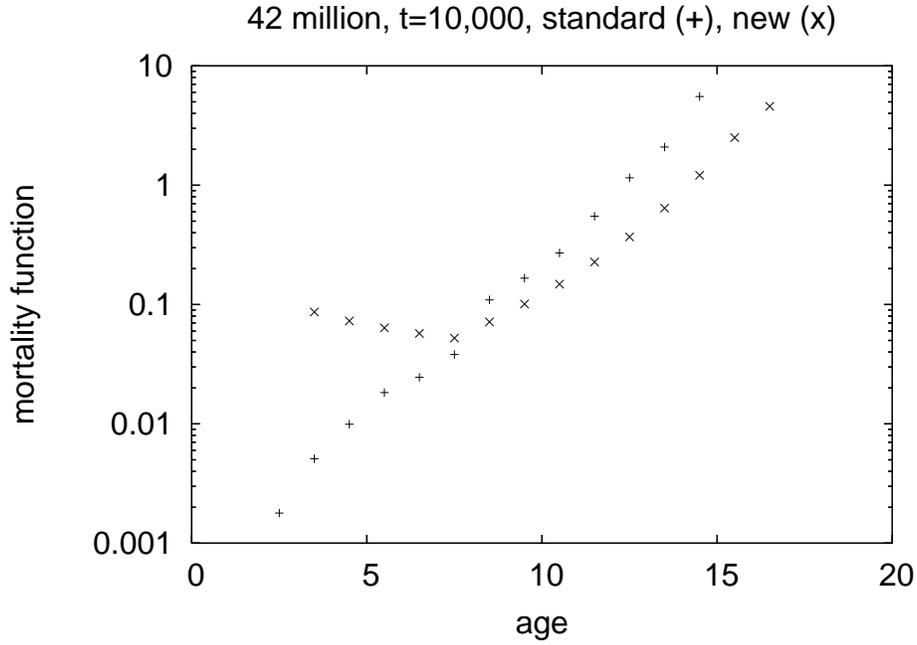}
\end{center}
\caption{Mortality function $\mu$ versus age $a$ in the usual Penna ageing
model (+) and in our modification (x).
}
\end{figure}

\end{document}